\documentstyle[prb,aps,psfig,floats]{revtex}

\begin{document}
\draft

\widetext
\title{Finite size Spin Wave theory of the triangular Heisenberg model}
\author{Adolfo E. Trumper} 
\address{Instituto de F\'{\i}sica Rosario (CONICET) and 
Universidad Nacional de Rosario, \\  Bv. 27 de febrero 210 bis,
(2000) Rosario, Argentina.}
          
\author{Luca Capriotti and Sandro Sorella}

\address{Scuola Internazionale Superiore di Studi Avanzati
         and Istituto Nazionale di Fisica della Materia (INFM),\\
         via Beirut 2-4, 34013 Trieste, Italy.} 
\date{\today}
\maketitle
\begin{abstract}
We present a finite size spin wave calculation on
the Heisenberg antiferromagnet on the triangular lattice 
focusing in particular on the low-energy part of the excitation 
spectrum. For $s=1/2$ the good agreement with the exact diagonalization
and quantum Monte Carlo results supports the reliability of the
spin wave expansion to describe the low-energy spin excitations
of the Heisenberg model even in presence of frustration. 
This indicates that the spin susceptibility of the
triangular antiferromagnet is very close to the
linear spin wave result.

\end{abstract}

\section{Introduction}

The antiferromagnetic Heisenberg model represents 
the  prototype 
of a quantum many body system where the Hamiltonian does not commute 
with the order parameter -the staggered magnetization-.
Therefore, quantum fluctuations are important and may play a 
crucial role in the ground state properties of the system. 
In principle they might destroy the long-range order  
of the classical minimum energy configuration depending, basically, 
on the dimensionality and the topology of the system.

In two dimensions and for a triangular lattice, quantum fluctuations are also
strongly enhanced by frustration so that Fazekas and Anderson \cite{anderson}
argued the possible stabilization of 
a resonating valence bond (RVB) ground state, i.e., a state which does not 
break the $SU(2)$ symmetry of the Hamiltonian.
However, after a long period of intensive theoretical and numerical investigations
\cite{laughlin,huse,singh,elstner,warman,jolicoeur,miyake,manuel,azaria,runge,bernu,chubukov},
no definite conclusion has been settled on 
the ground state properties of this system.
In fact -unlike the antiferromagnet on the square lattice where there
is a general consensus about the ordered nature of the ground state
even for $s=1/2$- in the frustrated cases the lack of exact 
analytical results is accompanied by difficulties in applying
stochastic numerical methods, as their reliability
is strongly limited by the well-known {\em sign problem}.
A few years ago Bernu {\em et al.} \cite{bernu}, with a deep analysis of 
the exact excitation spectra obtained for small clusters with the 
Lanczos technique, already evidenced an  antiferromagnetic 
$120^{\circ}$ ordering in the ground state. 
Only very recently, however, quantum 
Monte Carlo (QMC) calculations \cite{letter} have allowed to 
extrapolate to fairly large system sizes giving evidence 
of a quantum N\'eel order with the
order parameter $m^{\dagger}$ reduced by about 
$59\%$ from its classical value. 

Assuming as established the long-range N\'eel order of the ground state,
in this paper we address the reliability of the spin wave (SW) 
theory as an analytical tool to treat the  Heisenberg antiferromagnet 
on the triangular lattice. To this purpose, we generalize 
to the latter model a previously developed \cite{zhong}
SW theory specialized to finite size systems, and we study the low-lying
excited states. We will show that the agreement with numerical
results obtained by exact diagonalization and QMC 
is very good, thus confirming the effectiveness of SW 
theory in describing the ground state properties of the triangular 
Heisenberg model.     
 
\section{Spin wave calculation}

Several attempts to generalize SW theory to finite sizes can be found in the
literature \cite{zhong,takahashi,hirsch}. Here we will follow the 
method proposed in the mentioned
Ref.~\cite{zhong} which allows to deal with finite clusters avoiding
the spurious Goldstone modes divergences in a straightforward
way, and, in particular, without imposing any {\em ad hoc} holonomic
constraint on the sublattice magnetization.

Assuming the classical ${\bf Q}=(4\pi/3,0)$ 
magnetic structure lying in the $x{-}y$ plane and 
applying the unitary transformation which defines a
spatially varying coordinate system ($x^\prime{-}y^\prime{-}z^\prime$) 
in such a way that the $x^\prime$-axis coincides on each site 
with the local N\`eel direction, 
the transformed Heisenberg Hamiltonian reads:
\begin{eqnarray}
{\cal H}&=&J\sum_{\langle i,j \rangle} 
\Big[ \cos\left({\bf Q} \cdot ({\bf r}_j-{\bf r}_{i})\right) 
(S^{x^\prime}_i S^{x^\prime}_j+S^{y^\prime}_i S^{y^\prime}_j) 
\nonumber\\
&+& \sin\left({\bf Q}\cdot ({\bf r}_j-{\bf r}_i)\right)
(S^{x^\prime}_i S^{y^\prime}_j-S^{y^\prime}_i S^{x^\prime}_j)
+S^{ z^\prime}_i S^{z^\prime}_j \Big]
\label{heis}
\end{eqnarray}
where $J$ is the (positive) exchange constant between nearest 
neighbors, the indices $i,j$ label the points ${\bf r}_{i}$
and ${\bf r}_{j}$ on the $N$-site triangular lattice and
the quantum spin operators satisfy $|{\bf S}_{i}|^2=s(s+1)$. 
Then, using Holstein-Primakoff transformation for spin operators to order
$1/s$,
$S^{x^\prime}_{i} = s-a^{\dagger}_i a_i$,
$S^{y^\prime}_i = \sqrt{\frac{s}{2}}(a^{\dagger}_i+a_i)$,
$S^{z^\prime}_i = {\it i}\sqrt{\frac{s}{2}}(a^{\dagger}_i-a_i)$,
being $a$ and $a^{\dagger}$ the canonical creation and destruction Bose operators, after some algebra the Fourier transformed Hamiltonian results: 
\begin{eqnarray}
\label{HSW}
{\cal H}_{SW}&=&E_{cl} +
\nonumber  \\
&&3Js\sum_{\bf k}\Big[ A_{\bf k} a^{\dagger}_{\bf k}a_{\bf k}
+ \frac{1}{2}B_{\bf k}
(a^{\dagger}_{\bf k}a^{\dagger}_{-\bf k}+a_{\bf k}a_{-\bf k})\Big]
\end{eqnarray}
where $E_{cl}=-3Js^2N/2$ is the classical ground state energy, 
$A_{\bf k}=1+\gamma_{\bf k}/2$,
$B_{\bf k}=-3~\gamma_{\bf k}/2$, 
$\gamma_{\bf k} {=} \left[ \cos\,(k_x)\,+2\,\cos\,(k_x/2)
\cos\,(\sqrt3\,k_y/2)\right]/3$ and ${\bf k}$ is
a vector varying in the first Brillouin zone of the lattice.
The Hamiltonian
${\cal H}_{SW}$, can be diagonalized for ${\bf k} \neq 0, \pm {\bf Q}$ 
introducing the well-known Bogoliubov transformation, 
$a_{\bf k}=u_{\bf k}\alpha_{\bf k}+v_{\bf k}\alpha^{\dagger}_{-{\bf k}}$,
with
\begin{equation}
\label{transf}
u_{\bf k}=
\left(\frac{A_{\bf k}+
\epsilon_{\bf k}}{2\epsilon_{\bf k}}\right)^{1/2},
v_{\bf k}=-{\rm sgn}(B_{\bf k}) \left(\frac{A_{\bf k}-
\epsilon_{\bf k}}{2\epsilon_{\bf k}}\right)^{1/2}
\end{equation}
where $\epsilon_{\bf k}=\sqrt{A^2_{\bf k}-B^2_{\bf k}}$ is the  
SW dispersion relation. Such a diagonalization leads to 
\begin{eqnarray}
\label{diag}
{\cal H}_{SW}^0&=&E_{\it cl}+  
\frac{3Js}{2}\hspace{-2mm}\sum_{{\bf k} \neq 0, \pm {\bf Q}}
(\epsilon_{\bf k}-A_{\bf k})+
\nonumber  \\
&&\frac{3Js}{2}\hspace{-2mm}\sum_{{\bf k} \neq 0, \pm {\bf Q}}
\epsilon_{\bf k}
 (\alpha^{\dagger}_{\bf k}\alpha_{\bf k}~+
 \alpha^{\dagger}_{\bf -k}\alpha_{\bf -k}~).
\end{eqnarray}

The Goldstone modes, instead, cannot be diagonalized 
with this transformation since they become  
singular for ${\bf k}={\bf 0}$ and ${\bf k}= {\pm \bf Q}$. 
For infinite systems such modes 
do not contribute to the integrals in Eq.~(\ref{diag}), 
but in the finite case they are important and they must 
be treated separately. 
 By defining the following Hermitian operators 
\begin{eqnarray}
Q_x&=&\frac{{\it i}}{2}(a^{\dagger}_{\bf Q}
+a_{-{\bf Q}}-a_{\bf Q}-a^{\dagger}_{-{\bf Q}})~,\nonumber \\
Q_y&=&\frac{1}{2}(a^{\dagger}_{\bf Q}
+a_{-{\bf Q}}+a_{\bf Q}+a^{\dagger}_{-{\bf Q}})~,\nonumber \\
Q_z&=&i (a^{\dagger}_0-a_0) ~,
\end{eqnarray}
such that, $[Q_{\alpha},Q_{\beta}]=0$ and 
$[Q_{\alpha},{\cal H}_{SW}]=0$ for $\alpha,\beta =x,y,z$,
the contribution of the singular modes, ${\cal H}_{SM}$, in Eq.(\ref{HSW})
can be expressed in the form 
$$
{\cal H}_{SM}=-3JsA_{\bf 0}
+3Js\frac{A_{\bf 0}}{2}\left[Q_x^{ 2}+Q_y^{2}+Q_z^{2}\right].
$$
Then, taking into account the fact that to the leading 
order in $1/s$, $Q_{\alpha}=S^{\alpha}\sqrt{2/Ns}$,
where $S^{\alpha}$ are the components of the total spin, 
${\cal H}_{SM}$ may be also rewritten in the more physical form
$$
{\cal H}_{SM}=-3JsA_{\bf 0}
+3J\frac{A_{\bf 0}}{N}\left[(S^x)^2+(S^y)^2+(S^z)^2\right],
$$
which clearly favors a singlet ground state (for an even number of sites) 
being $A_{\bf 0}$ positive definite. 
This result is highly non
trivial since we have recovered the Lieb-Mattis \cite{lieb} property which 
has not been demonstrated for non bipartite lattices. 
Actually, a similar result is obtained by solving exactly
the three Fourier components ${\bf k}={\bf 0},\pm {\bf Q}$ of the 
Heisenberg model \cite{bernu}; however, our treatment allows to
construct  a formal expression for the SW ground state  
on finite triangular lattices which keeps the correct singlet behavior.
In fact, starting from the usual SW ground state, composed by
the $120^{\circ}$ classical N\'eel order plus the 
zero point quantum fluctuations (i.e,  zero Bogoliubov quasiparticles),
$$
|0\rangle=\prod_{{\bf k}\neq {\bf 0},\pm{\bf Q}} u^{-1}_{{\bf k}}
{\rm exp}{\Big[\frac{1}{2}\frac{v_{{\bf k}}}{u_{{\bf k}}} 
a^{\dagger}_{\bf k}a^{\dagger}_{-{\bf k}}\Big]}|N\rangle
$$
with $|N\rangle=\prod_i|S^{x^\prime}_i=s\rangle$, the corresponding
singlet wavefunction is obtained by projecting $|0\rangle$ onto 
the subspace $S=0$:
\begin{equation}
\label{project}
|\psi_{SW}\rangle=\int^{\infty}_{-\infty}\hspace{-1mm}d\alpha 
\int^{\infty}_{-\infty}\hspace{-1mm}d\beta\int^{\infty}_{-\infty}
\hspace{-1mm}d\gamma~
e^{{\it i}\alpha Q_x+{\it i}\beta Q_y+{\it i}\gamma Q_z}|0\rangle
\end{equation}
and reads
$|\psi_{SW}\rangle\sim e^{(-a^{\dagger}_{\bf Q}a^{\dagger}_{- \bf Q}+
\frac{1}{2}a^{\dagger}_{\bf 0}a^{\dagger}_{\bf 0})} |0\rangle~.$
In particular the singular modes have no contribution
to the ground state energy 
while the computation of the order parameter
(defined as in Ref.~\cite{bernu}), $m^{\dagger}$, 
requires their remotion:
\begin{equation}
m^{\dagger}=\sqrt{\langle (S^{x^\prime}_i)^2\rangle}=s-\frac{1}{N} \sum_{{\bf k}\neq 
{\bf 0},\pm {\bf Q}}v^2_{\bf k}~.
\end{equation}

Even for $s=1/2$, the previous SW calculation predicts a very good
quantitative agreement with exact results on small
clusters ($N \le 36$) of both ground state energy and sublattice
magnetization \cite{bernu}. 
For larger lattice sizes (see Table \ref{table1}), 
a comparison of the SW predictions 
with the available QMC results \cite{letter} 
also shows a surprisingly good agreement for the ground state energy 
which is conserved also with respect to the extrapolated values 
in the thermodynamic limit. 
Furthermore the QMC estimate
of the order parameter in the thermodynamic limit
$m^{\dagger}_{\rm QMC} = 0.41(1)$ can be 
reasonably compared with the SW prediction 
$m^{\dagger}_{\rm SW} = 0.478$ .
In the next sections we will also show that 
the low-energy spectra on finite sizes, calculated within the SW
approximation, compare very accurately
with the available numerical results. 
This agreement supports the numerical evidence for an ordered ground state in the present model.

\squeezetable

\begin{table}
\begin{tabular}{dcccc}
         $N$            &   36    &  48     &  108    & $\infty$ \\
\tableline
$E_{\rm SW}/JN $        &-0.5497   &-0.5464   &-0.5410   &-0.5388 \\
$E_{\rm QMC}/JN$        &-0.5581(1)&-0.5541(1)&-0.5482(1)&-0.5458(1)\\
\end{tabular}
\caption{Ground state energy
for the $s=1/2$ Heisenberg
antiferromagnet on the triangular lattice, as obtained
within the finite size SW theory and with
a QMC \protect\cite{letter} calculation. For $N=36$
the exact value of the energy
is  $E_{0}/JN=-0.5604$.
}
\label{table1}
\end{table}

\section{Low-energy spin wave spectrum}

In this  section, we show how to construct the low-lying energy
spectra $E(S)$  for finite systems. 
Following Ref.\cite{lavalle}, a magnetic field in the $z$-direction is 
added to stabilize  the desired total spin excitation $S$,
$$
{\cal H}^h_{SW}={\cal H}_{SW}-hs\sum_{i}S_{i}^{z}.
$$
Classically, the new solution is the $120^{\circ}$ N\'eel order canted by 
an angle $\theta$ along the direction of the field $h$. In order to develop a SW 
calculation, a new rotation around $y^\prime$-axis is performed on the spin
operators and it can be proven that ${\cal H}^h_{SW}$
takes the same form of Eq.(\ref{HSW}) with renormalized coefficients $A_{\bf k}$
and $B_{\bf k}$:
$$
A^h_{\bf k}=1+\gamma_{\bf k}\left[\frac{1}{2}-
\frac{3}{2}(\frac{2h}{z3J})^2\right]\:\:
B^h_{\bf k}=-\frac{3}{2}\gamma_{\bf k} \left[1-(\frac{2h}{z3J})^2\right]~,
$$
being $\frac{2h}{z3J}=\sin\theta$. In the present case the only singular mode is ${\bf k}={\bf 0}$ 
and its contribution is given by
$$
{\cal H}_{SM}=-\frac{3Js}{2}\;A^h_{{\bf 0}} 
+ 3J \frac{A^h_{{\bf 0}}}{N}\: (S^z-Ns\; \sin\theta)^2,
$$
which now  favors  a value of $S^{z}$ 
consistent  with the applied field, at the classical level. 
The Hellmann-Feynman theorem relates the latter quantities as it follows:
\begin{eqnarray}
\langle S^z_{i} \rangle&=&-\frac{1}{Ns}\frac{\partial}{\partial h}E(h)
\nonumber \\
&=&
s\frac{2h}{z3J} \left[1+\frac{1}{2Ns} \sum_{{\bf k}\neq0} \gamma_{\bf k}
\sqrt{\frac{A^h_{\bf k}+B^h_{\bf k}} {A^h_{\bf k}-B^h_{\bf k}}}\right]
\nonumber
\end{eqnarray}

\begin{figure}
\centerline{\psfig{bbllx=65pt,bblly=180pt,bburx=520pt,bbury=600pt,%
figure=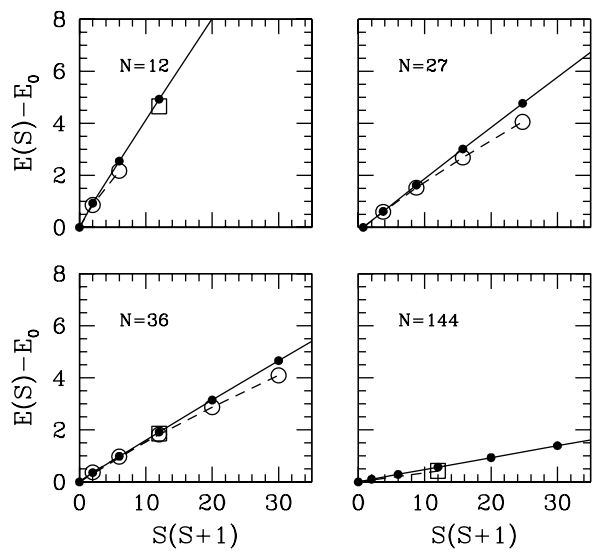,width=80mm,angle=0}}
\caption{SW (full dots and continuous line),  exact 
(empty dots and dashed lines)  and QMC \protect\cite{letter}
(empty squares) low-energy spectra as a function
of $|{\bf S}^2|=S(S+1)$ for $N=12,27,36,144$ and $s=1/2$. } 
\label{fig1}
\end{figure}

\noindent where 
\begin{equation}
\label{EH}
E(h)=E_{cl}-\frac{1}{2}(sh)^2\frac{2N}{3zJ}-3Js\frac{N}{2}+
\frac{3Js}{2}\sum_{\bf k}\epsilon^h_{\bf k}.
\end{equation}
and $\epsilon_{\bf k}^h=\sqrt{(A^h_{\bf k})^2-(B^h_{\bf k})^2}$.
In particular, the expansion to order $(sh)^2$ of the two first terms in 
Eq.(\ref{EH}) for $h\rightarrow0$ gives the classical perpendicular
susceptibility $\chi_{cl}=1/9J$, while taking the whole expression 
the known SW result \cite{chubukov} $\chi_{SW}/\chi_{cl}=1-0.449/2s$ 
is recovered. 
Finally, in order to evaluate 
the energy spectrum $E(S)$ of the $s=1/2$ case, 
a Legendre transformation $E(S)=E(h)+hsS$ has to be  performed.   
The advantage of this expression is that  
$E(S)$  can be computed within our framework 
for any size of the lattice and, in particular, in 
the thermodynamic limit. Of course our SW spectra are biased
by the underlying classical N\'eel order which is
invariant under the point group  $C_{3v}$
and under translations of the magnetic sublattices.

\begin{figure}
\centerline{\psfig{bbllx=65pt,bblly=160pt,bburx=505pt,bbury=690pt,%
figure=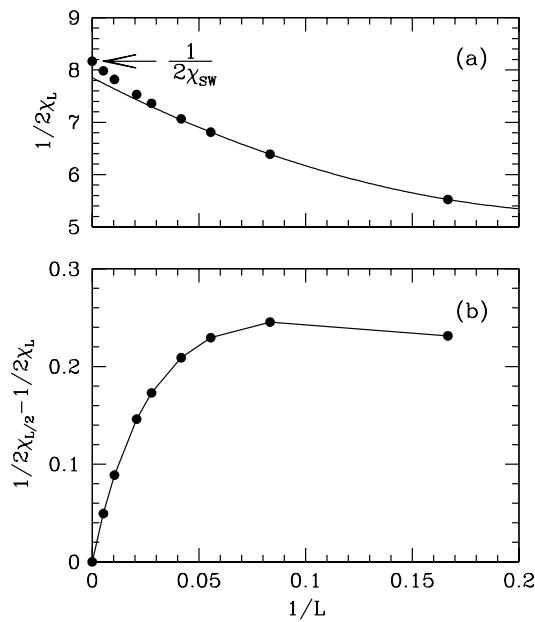,width=70mm,angle=0}}
\caption{Size dependence of 1/2$\chi_{L}$ (a) 
and of $1/2\chi_{L/2}-1/2\chi_{L}$ (b) obtained according to
Eq.~(\ref{inv}) using  the ($s=1/2$) SW excitation spectra.
The continuous line is a quadratic fit for $L<18$ in (a)
and a guide for the eye in (b).}
\label{fig2}
\end{figure}

\subsection{Comparison with numerical results}

The occurrence  of symmetry breaking
in the ground state  for $N\rightarrow \infty$ 
can be evidenced from the structure of the finite size
energy spectra. In particular, 
it is well known that when long-range order is present in
the thermodynamic limit, the low-lying excited states of
energy $E(S)$ and spin $S$ are predicted to behave as
the spectrum of a free quantum rotator (or {\em quantum top})
\begin{equation}
E(S) -E_0 \propto \frac{S(S+1)}{N}~,
\label{rotator}
\end{equation}
as long as $S \ll \sqrt{N}$.  
Fig.~\ref{fig1} shows $E(S)$ vs $S(S+1)$ calculated within the SW theory
compared with the exact \cite{bernu} (for $N\le36$) and QMC \cite{letter}
(for $S=3$, up to $N=144$) 
results, being the two last also consistent with  the above mentioned 
spatial symmetries of the classical  N\'eel state\cite{expla}. 
Remarkably SW theory turns out to be efficient 
to reproduce the low-energy spectrum in the whole range of sizes.
Furthermore, we can extend our calculation to the thermodynamic 
limit and observe easily the collapse of a macroscopic  number of
states with different $S$ to the ground state as 
$N\rightarrow\infty$. This clearly gives rise to a  broken $SU(2)$ 
symmetry ground state, as expected within the SW framework.

\subsection{Spin susceptibilities and anomalous finite size effects}

Whenever the quantum top law (\ref{rotator}) is verified, the quantity
\begin{equation}
\left[ 2 \chi_{S} \right]^{-1} = N E(S)\left[S(S+1)\right]^{-1}~, 
\label{inv}
\end{equation}
should approach the physical inverse susceptibility $1/2\chi_{\rm SW}$
for infinite size and for any spin excitation $S\ll N$. This feature is
clearly present in the SW theory and it is shown in Fig.~\ref{fig2}(a)
where the $1/2\chi_{S}$  is plotted for $S=L \equiv \sqrt{N}$ and approaches
the predicted value ($1/2\chi_{\rm SW}=8.167$), even if the correct asymptotic
scaling $1/2\chi_{L}=1/2\chi_{\rm SW}+ a/L + b/L^2$ turns out
to be satisfied only for very large sizes ($L\geq 36$). Such feature
is also shared by the Heisenberg antiferromagnet on the square lattice
where a similar SW analysis \cite{lavalle} has allowed to account for the anomalous 
finite size spectrum resulting from an accurate QMC calculation.
Furthermore, similarly to the latter case, a non-monotonic 
behavior of $1/2\chi_{L/2}-1/2\chi_{L}$ (Fig.~\ref{fig2}(b)), 
which should extrapolate to 0 as $1/L$ according to 
the quantum top law, 
persists also in presence of the frustration within the SW approximation
and is likely to be a genuine feature of the Heisenberg model. 

\section{Concluding Remarks}
\label{tower}

In conclusion, we have applied a previously developed finite size spin wave 
theory to the Heisenberg model on the triangular lattice. 
Comparison for the case $s=1/2$ 
of the low-energy part of the spin wave 
spectra  with the exact and the more recent quantum Monte Carlo results 
reveals a very good agreement.

The accuracy of our results on finite sizes indicates that 
spin wave theory is a reliable analytical approximation to describe 
the ground state properties of the present model in the thermodynamic
limit $N\to\infty$.
In particular the effectiveness of the spin wave theory
in reproducing the finite size spectrum  
strongly suggests that the value of the spin susceptibility
should be very close to the spin wave prediction. This represents the main finding
of this paper.
Furthermore we have found, to order 1/s, anomalous
finite size effects for the spin susceptibilities for $s=1/2$,
similar to the case of the square lattice.

Finally, this 
study provides further support to  the recent numerical evidences
about the existence of long-range N\'eel order in the ground state 
of the present frustrated model.

\acknowledgements

This work was supported in part by INFM (PRA HTCS), MURST (COFIN97)
and Fundaci\'on Antorchas (A.E.T.).
It is a pleasure to acknowledge stimulating discussions with
C. Lhuillier, M. Capone, M. Calandra, F. Becca, A. Parola, 
G. Santoro, V. Tognetti and L. O. Manuel.


\begin{references}
\bibitem{expla} In order to perform a faithful comparison, we have computed 
the spin wave  energy $E(S)$
using the same clusters  of the numerical results, i.e., with
periodic boundary conditions
and all the spatial symmetries of the
infinite lattice  that do not frustrate the classical
N\'eel order (see for details Ref.\cite{bernu}).
\end{references}
\end{document}